\documentclass[11pt]{article}
\usepackage{moriond,epsfig}

\begin{document}
\vspace*{4cm}
\title{INDIRECT DARK MATTER SEARCH WITH AMS-02}

\author{ S. DI FALCO }

\address{on behalf of the AMS Collaboration\\
INFN \& Universit\`a di Pisa, Largo B. Pontecorvo 3,\\
56100 Pisa, Italy}

\maketitle\abstracts{
The Alpha Magnetic Spectrometer (AMS), to be installed on the International
Space Station, will provide data on cosmic radiations in the energy range from 
0.5 GeV to 3 TeV. 
The main physics goals are the anti-matter and the dark matter searches. 
Observations and cosmology indicate that the Universe may include a large 
amount of unknown Dark Matter. 
It should be composed of non baryonic Weakly Interacting Massive Particles 
(WIMP). 
In R-parity conserving models a good WIMP candidate is the lightest SUSY 
particle. 
AMS offers a unique opportunity to study simultaneously SUSY dark matter in 
three decay channels resulting from the neutralino annihilation: 
$e^+$, antiproton and gamma. 
Either in the SUSY frame and in alternative scenarios (like extra-dimensions)
the expected flux sensitivities as a function of energy in 3 year exposure 
for the $e^+/e^-$ ratio, gamma and antiproton yields are presented.}

\section{The search for dark matter}
The evidence for dark matter existence has become more and more robust in the recent years: several independent indications were provided by rotation curves and mass to light ratios of galaxies, by X rays emissions from clusters and by  gravitational lensing. 
The very recent results from the WMAP collaboration \cite{wmap3} confirm that baryon matter density ($\Omega _b = 0.0223^{+0.0007}_{-0.0009}$) is largely insufficient to saturate the total matter density 
($\Omega _m=0.127^{+0.007}_{-0.013}$).

The known non baryonic dark matter candidates, such as massive neutrinos or massive black holes, don't reach the required dark matter density. Other, still undiscovered, possible candidates are the axions and the weakly interacting particles (WIMPs).
These latest include  as most favourites the {\em neutralino},
the lightest supersymmetric particle (LSP) \cite{jungman}, 
and the lightest Kaluza-Klein particle (LKP)
of certain extra-dimensions models \cite{servant}.

The search for WIMPs can be performed either {\em directly} or {\em indirectly}.
Direct obervations look for the nuclear recoils in the elastic scattering on nuclei  \cite{serfass}. 
Indirect  searches look for anomalies in the expected spectra of primary cosmic rays due to the annihilation of dark matter candidates.

Hints for anomalies in the positron and photon spectra around 10 GeV have been observed by the HEAT \cite{heat} and EGRET \cite{egret} collaborations, respectively.

The AMS02 experiment can confirm or disprove these hints by measuring simultaneously the spectra of positrons, photons and antiprotons. 
Together with the search for dark matter, the main goals of the experiment will be the search for antimatter and a precise measurement of cosmic ray fluxes in the energy range between 0.5 GeV and 3 TeV, including nuclei up to Z=26 and gamma rays.
\begin{figure} [htb]
\begin{center}
 \centerline{\mbox{\epsfig{file=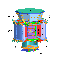,height=6 cm}}
\mbox{\epsfig{file=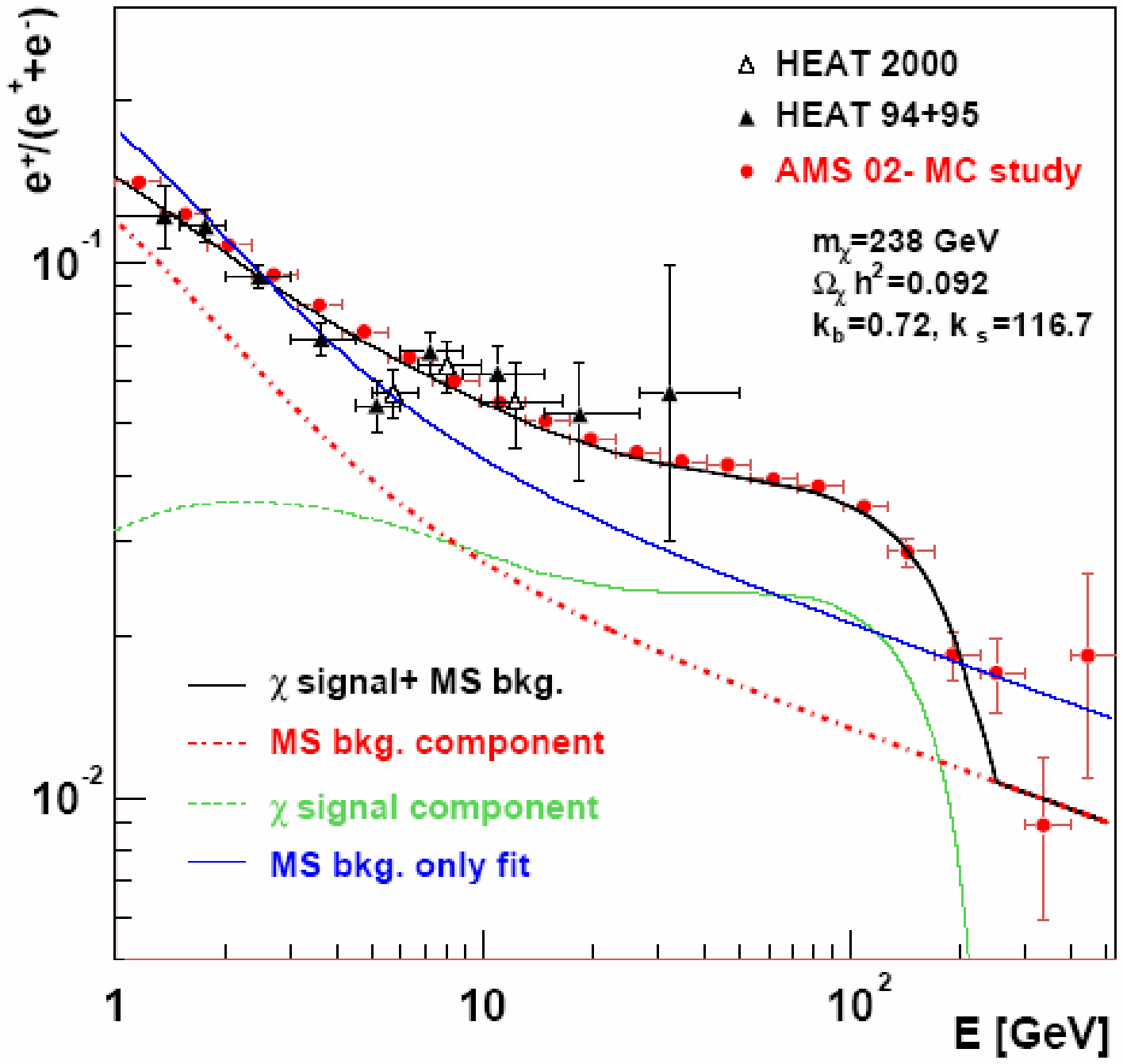,height=6 cm}}}
 \caption{\footnotesize a) The AMS-02 experiment layout. 
b) Example of the statistical accuracy on positron fraction measurement in 3 years in case of neutralino annihilation ($m_\chi$=238 GeV, boost factor =166). }
 \label{ams02flux}
\end{center}
\end{figure}

\section{The AMS experiment}
The Alpha Magnetic Spectrometer (AMS) \footnote{
A precursor flight (AMS-01) succesfully flew for 10 days in June 1998 on the Space Shuttle Discovery \cite{ams1}.}
is a cosmic ray detector which 
will operate on the International Space Station (ISS) for  at least three years. 
AMS will be ready for launch in 2008.

The experimental layout (fig.\ref{ams02flux}a) consists of a
{\em Transition Radiation Detector (TRD)} \cite{trd}, a {\em Time of Flight (TOF)} \cite{tof}, a {\em Silicon tracker (TRACKER)} \cite{tracker}, a {\em Ring Imaging Cherenkov (RICH)} \cite{rich}, an {\em Electromagnetic Calorimeter (ECAL)} \cite{ecal}, an {\em Anti-Coincidence Counter (ACC)} and a Star Tracker.
A large  superconducting magnet, the first operating on Space, will provide a bending power of $BL^2 = 0.85\ Tm^2$.

\section{Dark Matter signal in $e^+$ spectrum}

The measurement of positron spectrum requires all the AMS subdetectors:
a rejection factor of $10^2\div10^3$ on protons is obtained by TRD for proton energies up to 300 GeV \footnote{No transition radiation is expected for relativistic factors $\gamma<10^3$.};  
the time measurements by TOF and the track curvature by TRACKER permit the charge sign determination, so rejecting electrons up to 2 TeV; 
the Z measurement by TOF, TRACKER and RICH make the He background negligible; 
the velocity ($\beta$) measurement by RICH allows the rejection of protons up to 10 GeV; 
the energy deposit profile inside ECAL ensures a lepton/proton rejection factor of   $\sim10^3$; 
finally, the matching between TRACKER momentum and ECAL energy furtherly suppresses the hadronic component. 

Combining all these informations together a global rejection factor of $10^5$ for $p$ and $10^4$ for $e^-$ is achieved. 
The mean acceptance for positrons in the energy range form  3 to 300 GeV is 0.045 $m^2 sr$, with a proton contamination of $\sim 4\%$ \cite{maestro}.

Fig. \ref{ams02flux}b shows an example of {\em positron fraction} \footnote{This ratio is  preferred to the simple spectrum since less dependent on cosmic ray propagation.} measured by AMS if the excess on the HEAT data were due to the annihilation of 238 GeV neutralinos: a signal boost factor \footnote{These boost factors can be explained assuming a clumpy dark matter halo.} of $\sim 100$ has been used to fit the HEAT data \cite{baltz}. Significantly lower boost factors are required if the anomaly is due to  LKPs with masses of few hundreds GeV \cite{pochon}.

\begin{figure} [t]
\begin{center}
 \centerline{\mbox{\epsfig{file=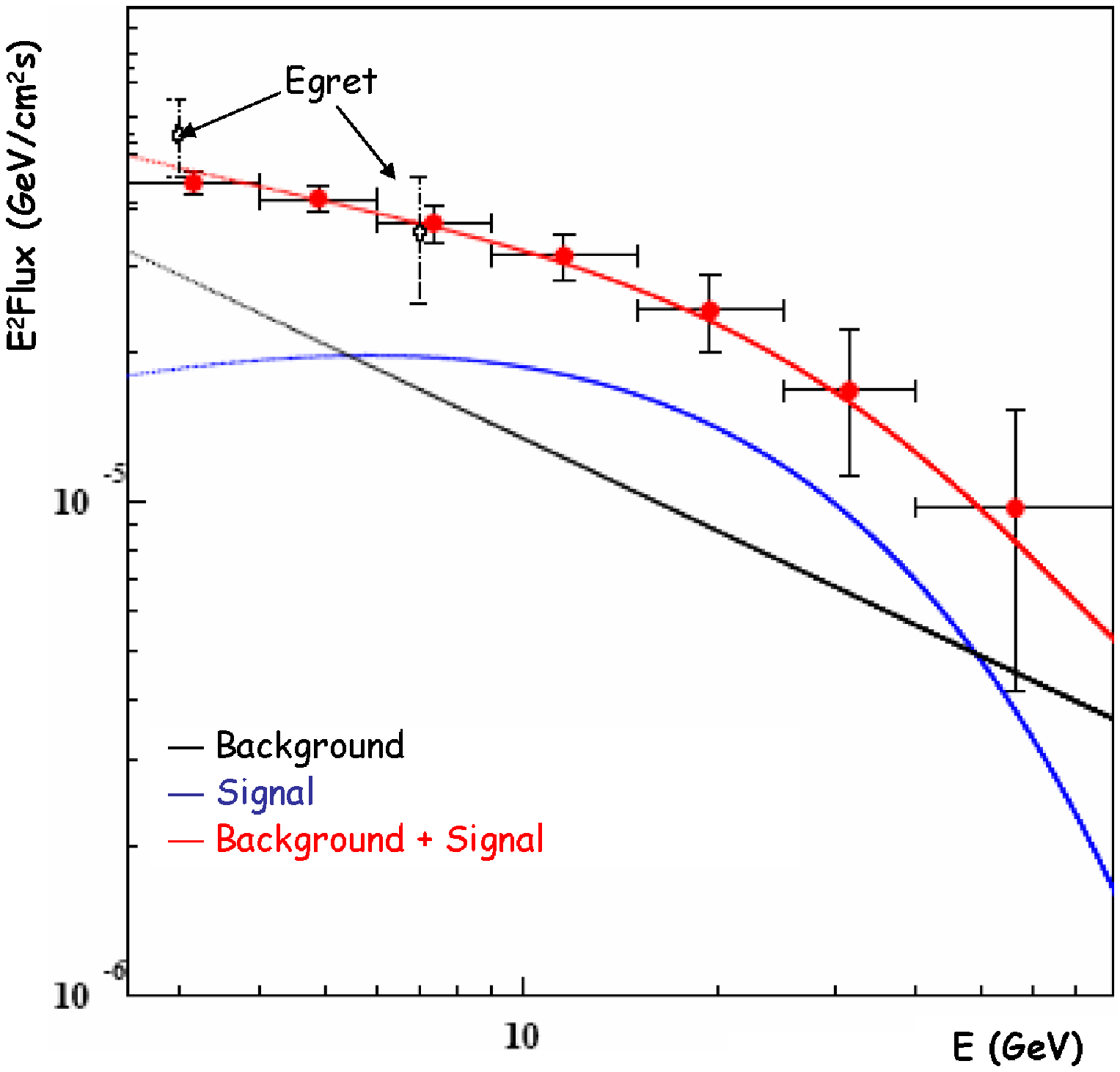,height=6 cm}}
\mbox{\epsfig{file=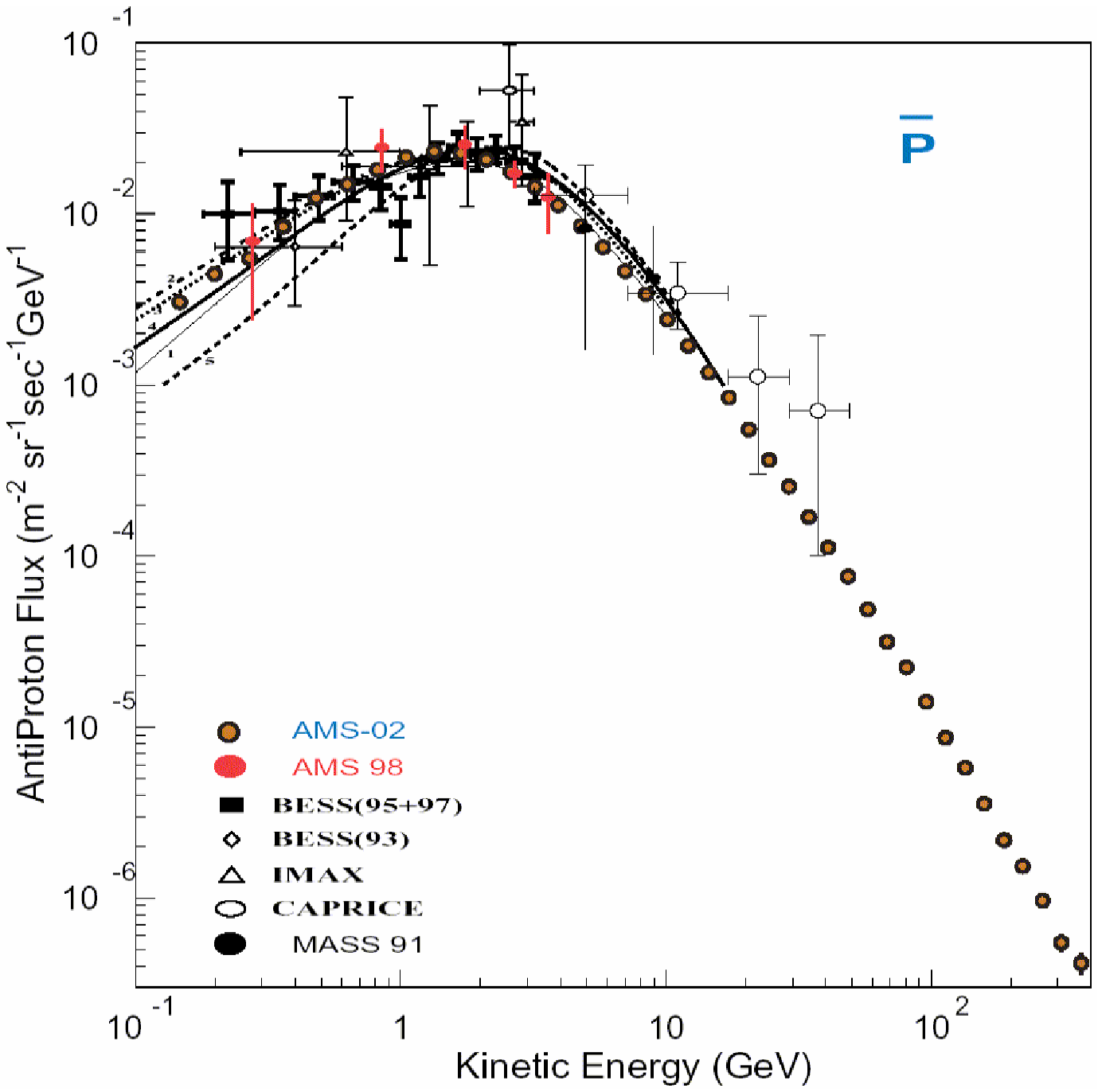,height=6 cm}}}
 \caption{\footnotesize a) Statistical accuracy in photon spectrum with 1 year data taking supposing a 208 GeV neutralino annihilation. 
b) Expected precision on the antiproton spectrum measurement by AMS in 3 years. } 
 \label{photprot}
\end{center}
\end{figure}

\section{Dark Matter signal in $\gamma$ spectrum}

In AMS photons are detected using two complementary techniques:
{\em photon conversions} in $e^+ e^-$ pairs that can be detected by the TRACKER and {\em direct photon} measurements in ECAL.

The conversion mode ensures an excellent photon angle resolution (0.02$^o$ at 100 GeV), a good energy resolution (3\% up to 20 GeV, 6\% at 100 GeV), a good acceptance (0.06 $sr\ m^2$ at 100 GeV) and a large field of view ($\sim43^o$).
The background is mainly due to $\delta$ rays and can be strongly reduced (rejection factor $\sim10^5$) vetoing with the TRD and cutting on the pair invariant mass.

The measurement with ECAL has an angular resolution of 1$^o$ at 100 GeV, an excellent energy resolution (3\% at 100 GeV), a large acceptance at high energies ($\sim 0.1\  sr\ m^2$ above 100 GeV) but a reduced field of view  of $\sim 23^o$.
The main background is due to $\pi ^0$s produced by proton interactions in the material surrounding ECAL. 
A rejection factor of $\sim 5\cdot 10^6$ on protons is obtained analyzing the cascade profile in ECAL and vetoing on charged particles \cite{pilo}. 

In 3 years the exposure to the galactic center \footnote{This quantity is not important for positrons or antiprotons since they suffer large energy losses and scatterings so that they lose memory of their initial direction.} will amount to 40 days for the conversion mode and to 15 days for the direct photon mode; 
the integrated acceptance will be practically the same for the two methods: 
$\sim1.5\cdot 10^5\ m^2 s$.

In the case of a photon spectrum anomaly due a 208 GeV neutralino annihilation fitting the EGRET data, the statistical evidence obtainable by AMS in 1 year of direct photon detection is shown in fig. \ref{photprot}a \cite{girard}.

In absence of any anomaly in the spectrum a large part of SUSY and Kaluza-Klein model parameter space could be excluded, in particular in the case of a cuspy halo profile \cite{agnieska}.

\section{Dark Matter signal in $\bar{p}$ spectrum}

In AMS the antiprotons will be detected  as negative single charged tracks 
reconstructed by TRD and TRACKER.
The acceptance for this signal is $\sim0.16\ sr\ m^2$ between 1 and 16 GeV and $\sim0.033  \ sr\ m^2$ between 16 and 300 GeV \cite{choutko}.
Misreconstructed proton interactions and unrecognized electrons are the main background sources: the rejection factors are better than $10^6$ against protons and $10^3 \div 10^4$ against electrons.

Fig. \ref{photprot}b shows the expected profile after 3 years together with the existing measurements of the antiproton flux: the dot dimensions correspond to the statistical error. 
The low energy part of the spectrum is consistent with the secondary production by interaction of primary cosmic radiation with the interstellar medium. 
The most interesting region is the one between 10 and 300 GeV where, supposing that large boost factors ($\sim10^3$) enhance the process, 
a dark matter annihilation could be revealed \cite{ullio}.

\section{Conclusions}
In three years of data taking the AMS experiment will be able to measure with unprecedented accuracy the rates and spectra of positrons, photons amd antiprotons in the GeV-TeV range, looking for an excess of events consistent with a dark matter annihilation signal.
Several models for dark matter candidates will be constrained by the new AMS data.

The simultaneous measurements of other fundamental quantities, such as the proton and electron spectra and the B/C ratio, will help to refine the astrophysical predictions enhancing the compelling evidence for a dark matter signal. 


\end{document}